# THE ECLIPSING BINARY SYSTEM SW LACERTAE: HOTOMETRIC STUDY

**Ahmed-Essam, A. El-Bassuny Alawy, and M. A. Hamdy**

**ABSTRACT:** *Blue and Visual photoelectric light curves have been achieved for the eclipsing binary SW Lac in 1995. Nine moments of minimum light were derived. Period of 0.320854 days have been determined as the mean of those derived from blue and visual observations. B and V light curves have been analyzed using W-D code where the system has been viewed through the ellipsoidal model. The system geometrical and physical parameters were determined and compared with those published recently.*

## INTRODUCTION

The eclipsing binary SW Lacertae (BD=+37° 4717, HD=216598, SAO=72820, HIC=113052) is one of the most often observed W UMa-type systems due to its variable light curve and period variation. Since the discovery of SW Lac variability by Miss Ashall (as reported by Leavitt, 1918) on plates taken at Harvard observatory, numerous photometric as well as few spectroscopic observations for this system have been done. It is a good candidate of W type W UMa systems exhibiting:
1. Short period (about 0.3 day) of the light cycle.
2. Partial eclipses while the primary one is an occultation one.
3. Both variable period and light curve.

In addition to these characteristics, the system posses the largest mass ratio (q=0.88) ever found among contact eclipsing binaries.

## PRESENT OBSERVATIONS

Photometric observations of the eclipsing binary system SW Lac. have been obtained through 5 nights during the period from 18 to 24 September 1995 using the 74-inch Kottamia reflector of NRIAG, Egypt. A single beam photoelectric photometer equipped with an EMI9862 photomultiplier cooled thermoelectrically. The measurements were carried out through B and V wide pass-band filters; closely match those of the standard Jonson system (Jonson, 1953). The star BD = +37° 4715

_________________________________________________________________
*National Research Institute of Astronomy and Geophysics, Helwan, Cairo, Egypt*

has been served as a reference star since it has been found to be nonvariable by many investigators (e.g. Rucinski, 1968; Faulkner and Bookmyer, 1980; and Niarchos, 1987). Table (1) lists the observational data of both systems SW Lac., moreover, the comparisons star.

Table 1: Observational Data of the System SW Lac.

| Star Name | BD No. | R. A. 2000 h m s | Dec. 2000 ° ′ ″ | Mag. | Spec. Type | No. of Obs. B V |
|---|---|---|---|---|---|---|
| SW Lac. | 37° 4717 | 22 53 41.1 | 37 56 17.7 | 8.5-9.4 | $K_{0v}$ | 1425  1425 |
| Comp. | 37° 4715 | 22 53 11.6 | 37 47 14.2 | 10.2 |  | 101   101 |

Measurements of the comparison star and sky were interspersed throughout the observing periods, with rather less frequent sky measures. These measures (vs. time) were fitted to polynomials, which were used to correct for both changing extinction (comparison data) and sky contribution in the variable observations. The heliocentric correction was applied to all observations. A total of 1425 observations were achieved for the variable in each filter. Five light curves for the primary and secondary eclipses were obtained and plotted in Figures 1 to 5.

## EPOCHS OF PHOTOMETRIC MINIMA

New nine times of minima of SW Lac. (Five primaries and four secondaries) were derived from the present photometry. The moments of these minima were calculated using the software package AVE (Barbera, 1996), that employs the method of kwee & Woerden (1956). The deduced times of minima are listed in Table (2) together with the probable error (P.E.), type of minima (Min.), the band of observations used (Filter), besides their O-C's residuals (O-C), and cycle number (E) with respect to the ephemeris given by Niarchos (1987).

$$\text{Hel. JD. (Min.I)} = 2444499.5264 + 0.3207204*E \quad (1)$$

No significant difference (within error quoted) was found between blue and visual times of minima implying that it is not a function of colour. Hence, their weighted means can be used for phasing the nightly light curve and period behaviour study.

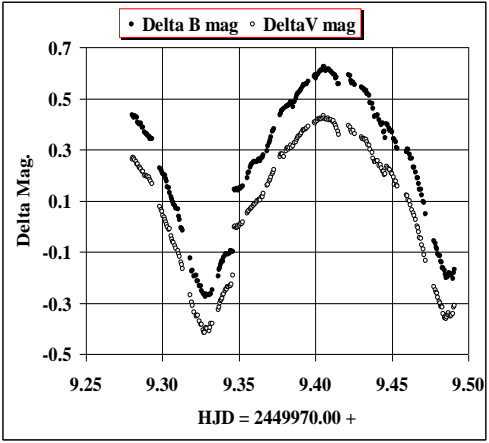

**Fig. 1: Delta B and V Observations of SW Lac at 18 Sep. 1995**

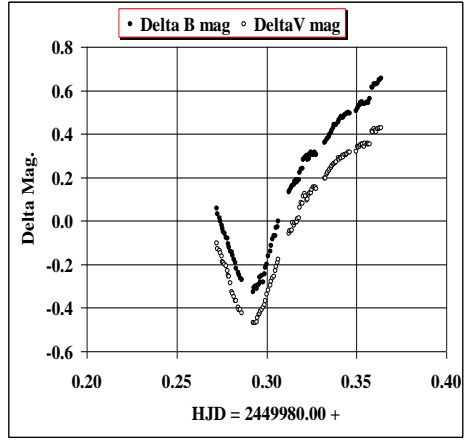

**Fig. 2: Delta B and V Observations of SW Lac at 19 Sep. 1995**

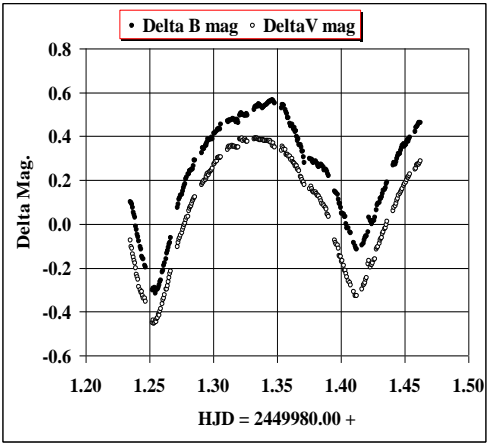

**Fig. 3: Delta B and V Observations of SW Lac at 20 Sep. 1995**

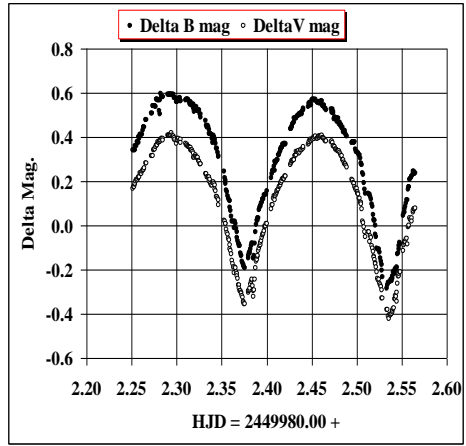

**Fig. 4: Delta B and V Observations of SW Lac at 21 Sep. 1995**

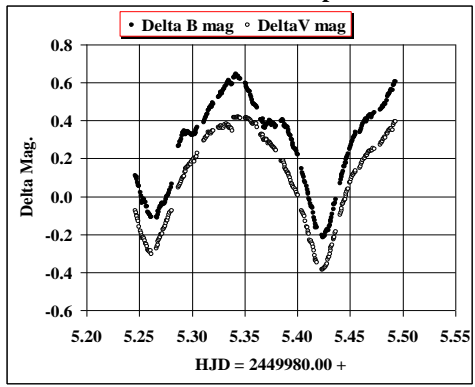

**Fig. 5: Delta B and V Observations of SW Lac at 24 Sep. 1995**

Table 2: Epochs Minimum Light of SW Lac.

| Date | HJD=2449900+ | P.E. | Min. | Filter | O-C | E |
|---|---|---|---|---|---|---|
| 18 Sep. 1995 | 79.328706 | 0.000440 | I | B | -0.03186 | 17086.0 |
|  | 79.328877 | 0.000185 | I | V | -0.03187 | 17086.0 |
|  | 79.486149 | 0.000160 | II | B | -0.02951 | 17086.5 |
|  | 79.485825 | 0.000464 | II | V | -0.02949 | 17086.5 |
| 19 Sep. 1995 | 80.291685 | 0.000307 | I | B | -0.02552 | 17089.0 |
|  | 80.291889 | 0.000464 | I | V | -0.02554 | 17089.0 |
| 20 Sep. 1995 | 81.254498 | 0.000089 | I | B | -0.02528 | 17092.0 |
|  | 81.253925 | 0.000025 | I | V | -0.02526 | 17092.0 |
|  | 81.413373 | 0.000165 | II | B | -0.02694 | 17092.5 |
|  | 81.412509 | 0.000167 | II | V | -0.02692 | 17092.5 |
| 21 Sep. 1995 | 82.533361 | 0.000141 | I | B | -0.02766 | 17096.0 |
|  | 82.535965 | 0.000123 | I | V | -0.02762 | 17096.0 |
|  | 82.375329 | 0.000055 | II | B | -0.02650 | 17095.5 |
|  | 82.375681 | 0.000150 | II | V | -0.02652 | 17095.5 |
| 24 Sep. 1995 | 85.424484 | 0.000141 | I | B | -0.02474 | 17105.0 |
|  | 85.423703 | 0.000098 | I | V | -0.02472 | 17105.0 |
|  | 85.261091 | 0.000220 | II | B | -0.02718 | 17104.5 |
|  | 85.261536 | 0.000238 | II | V | -0.02715 | 17104.5 |

## PHOTOMETRIC PERIOD DETERMINATION

The photometric orbital period of the system has been derived via AVE package (Barbera, 1996) where Fourier series have been applied. Only the observations achieved at 20 and 21 Sep. 1995 were combined and Fourier analyzed. This is due to of two reasons. First, both minima and maxima were observed; i.e. all characteristic phases are well represented. Secondly, the observations have been obtained in short time span, two consecutive nights. This may minimize errors resulted from light curve variation (a common phenomenon shown by most of late W UMa systems). A period of 0.320854 days (±0.000134) as the mean of those derived from blue and visual observations while the amplitudes deduced are 0.685 and 0.681 magnitude which are in good agreement with the light curve amplitude observed. Using the mean time of primary minima of the fourth night (21 Sep. 1995) and the new period, we obtain the new ephemeris to phase our observations as follows:

$$\text{Hel. J.D. (Min.I)} = 2449982.5347 + 0.320854 * E \quad (2)$$

We applied this ephemeris with the 4$^{th}$ night of observations to obtain and represent in Figure (6) and (7) the normalized points of complete light curve in B, and V filter in differential magnitude. From Figure (6) and (7), we can find the extrema of the system as tabulated in Table (3) and employ it for light curve modeling.

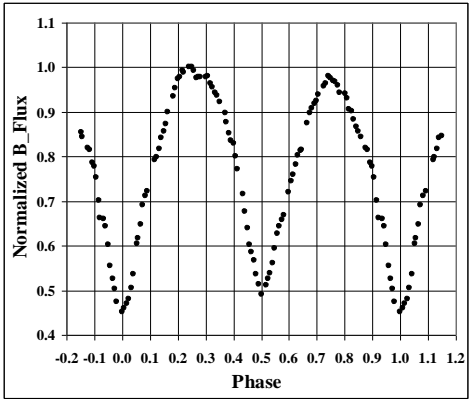

Fig. 6: Normalized points of B. Mag for SW Lac at 21 Sep. 1955

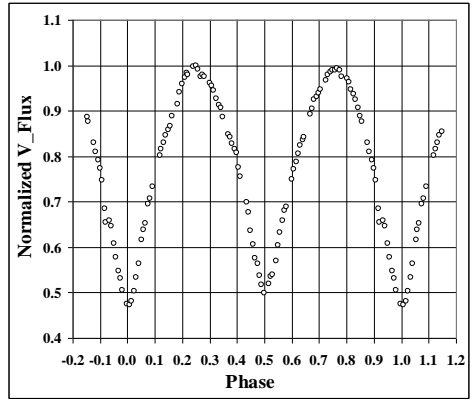

Fig. 7: Normalized points of V. Mag for SW Lac at 21 Sep. 1955

**Table 3: Extrema of SW Lac in the present work**

| Filter | Min.I-Max.I | Min.II-Max.II | Min.I-Min.II | Max.II-Max.I |
|---|---|---|---|---|
| B | 0.871 mag. | 0.761 mag. | 0.076 mag. | -0.034 mag. |
| V | 0.864 mag. | 0.704 mag. | 0.107 mag. | -0.053 mag. |

## GEOMETRICAL AND PHYSICAL PARAMETERS

Fortunately, the complete B and V light curves of SW Lac in two filters were represented by the observations of one night (21 Sep. 1995, see Figures 6 & 7). These observations with the new ephemeris (see equ. 2) have been used to analyze the light curves employing the recent version of Wilson-Devinney program (W-D, 2003). The model has been described and quantified in papers by Wilson & Devinney (1971), Wilson (1979, 1990, 1993), and Van Hamme & Wilson (2003).

Most of the previous literature (e.g. Leung et al. 1984 & Jeong et. al, 1994) showed that the system of SW Lac is overcontact and belongs to the W-Type of W UMa systems. Therefore, we choose mode 3 to carry out the photometric solution of the present light curves. We tried to fit the observed light curve by using the first part of WD code (LC program) under the conditions of the mode adopted in which $T_2$ is free parameter.

Some constraints are applied, the first is that the surface potential of star 2 is equal to that of star 1 ($\Omega_2 = \Omega_1$). Other constraints are that the gravity brightening, $g_i$, and bolometric albedo, $A_i$, parameters of star 2 are the same as those of star 1. The luminosity of star two, $L_2$, can be computed either from black body or from stellar atmosphere radiation formulas. We used the initial parameters deduced from previous subsection and those found by Jeong, et al. (1994), who assumed one spot on star 1.

We tried to fit the observed V-light curve through four models of parameters, differ only in spot assumptions, while the other parameters *(Inclination (i), Mass ratio (q), Bolometric albedo ($A_1$ & $A_2$), Surface potential ($\Omega_1$ & $\Omega_2$), Gravity exponents ($\alpha_1$ & $\alpha_2$), Surface temp. of both stars ($T_1$ & $T_2$)* are the same.
Model number 1 has no spots on both components of the system. Model number 2 has only one spot on the surface of star 1 (cool star). Model numbers 3 only one spot assumed on the surface of star 2 (hot star). While model number 4 has two spots, one on each star. The resultant fits curves throw the four models reveal that the best fit with the observed V light curve is the one of model number 2.
Applied the second part of WD code (DC program), with the model 2 parameters (which have only one spot on the surface of cool star) through the parameters' initial values, those wine been deduced from the last run of LC program. The temperature of the both stars $T_1$ and $T_2$, The phase shift; orbital inclination, surface potential, mass ratio, and spot parameters were adjusted. The other parameters were fixed as shown in Table (4). The final fit of the observations was plotted in Figures (8) and (9) for V and B filter respectively. The parameters derived have been tabulated in Table (4) with their standard deviation. The fit is good enough in the V and B curves except near the secondary minimum in V filter. There is a deviation between computed values and observations; the depth of the secondary minimum in observation is deeper than the computed one. This may be due to observational scatter. Using the geometrical and physical parameters in Table (4) with the Binary Maker 3.0 program (Bradstreet 2004) to produce the Roche geometry of the system in Figure (10) to show the degree of contact. The same parameters have been employed with the same program to display the system configuration at different phases (0.0, 0.25, 0.5, and 0.75) with the positions of the spots in Figure (11).

**Table 4: Adjusted Parameters of SW Lacertae for model No. 3 in V and B filters**

| Parameter | V filter | B filter |
|---|---|---|
| **Phase Shift** | 00.5026 ± 0.0005 | 00.5044 ± 0.0006 |
| Inclination (i) | 80°.817 ± 0°.257 | 80°.948 ± 0°.332 |
| Surface Temp. $T_1$ | 5379 °k ± 7.048 °k | 5371 °k ± 6.995 °k |
| Surface Temp. $T_2$ | 5521 °k ± 6.960 °k | 5529 °k ± 6.619 °k |
| Surface potential ($\Omega_1 = \Omega_2$) | 3.90644 ± 0.01196 | 3.91901 ± 0.01494 |
| Bolometric albedo ($A_1=A_2$) | 0.500 (fixed) | 0.500 (fixed) |
| Gravity exponents ($\alpha_1= \alpha_2$) | 0.320 (fixed) | 0.320 (fixed) |
| $L_1 / (L_1+ L_2)$ | 0.4146 (fixed) | 0.4033 (fixed) |
| $L_2 / (L_1+ L_2)$ | 0.5854 (calculated) | 0.5967 (calculated) |
| Mass Ratio (q) | 1.24812 ± 0.01044 | 1.21735± 0.012699 |
| Limb Darkening $x_1$ | 0.664 (fixed) | 0.803 (fixed) |
| Limb Darkening $x_2$ | 0.638 (fixed) | 0.777 (fixed) |
| Spot Latitude of cool star | 45°.00 (fixed) | 45°.00 (fixed) |
| Spot Longitude of cool star | 279°.53 ± 51°.30 | 346°.54 ± 33°.27 |
| Spot Radius of cool star | 17°.23 ± 1°.54 | 20°.35 ± 1°.97 |
| ($T_{spot} / T_{star}$) of cool star | 0.7500 (fixed) | 0.7500 (fixed) |
| % overcontact | 37 % | 31 % |
| $r_1$(pole) | 0.3605 ± 0.0019 | 0.3605 ± 0.0024 |
| $r_2$(pole) | 0.3974 ± 0.0017 | 0.3974 ± 0.0021 |
| $r_1$(side) | 0.3819 ± 0.0025 | 0.3819 ± 0.0031 |
| $r_2$(side) | 0.4234 ± 0.0023 | 0.4234 ± 0.0029 |
| $r_1$(back) | 0.4287 ± 0.0046 | 0.4287 ± 0.0056 |
| $r_2$(back) | 0.4659 ± 0.0038 | 0.4659 ± 0.0047 |
| $M_1$ & $M_2$ (Masses in Solar Units) | 1.220 & 0.978 | 1.207 & 0.991 |
| $R_1$ & $R_2$ (Radius in Solar Units) | 1.12 & 1.02 | 1.09 & 1.00 |
| $\Sigma \omega (O-C)^2$ | 0.03199 | 0.0424 |

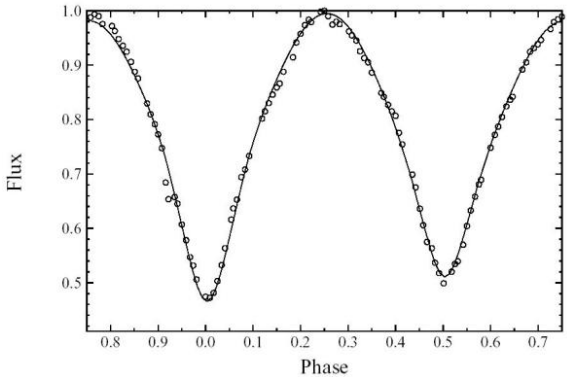

**Fig. 8: V light Curve of SW Lac, with the Fit Curve (Solid Line)**

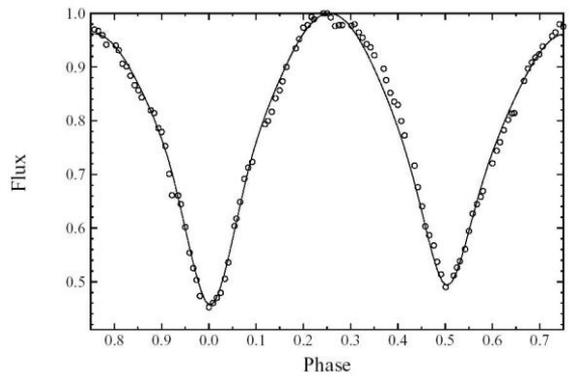

**Fig. 9: B light Curve of SW Lac, with the Fit Curve (Solid Line)**

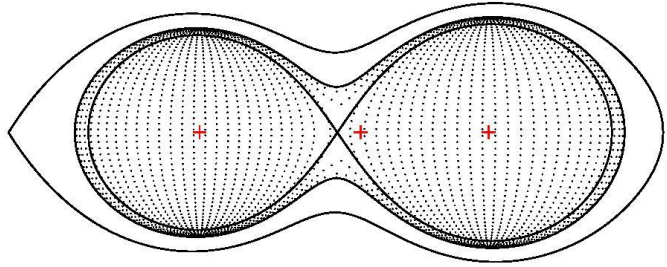

**Fig. 10: Roche Geometry of the system SW Lac.**

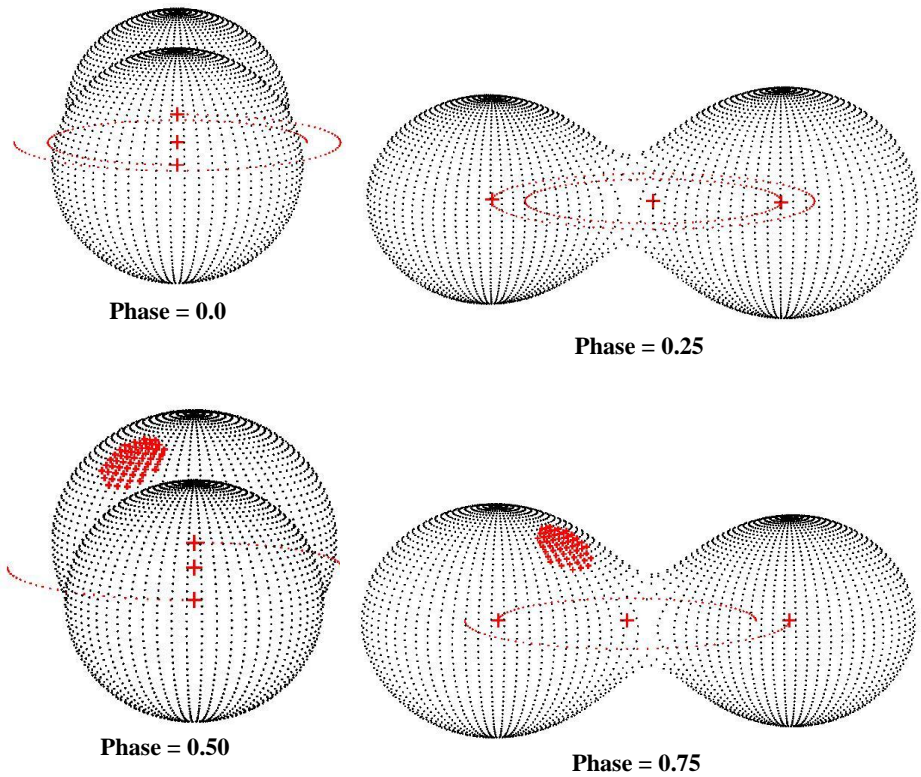

**Fig. 11: Shape of the system SW Lac. at phases 0.0, 0.25, 0.50, and 0.75**

## DISCUSSION AND CONCLUSION

The present photometry shows pronounced asymmetric light curves regarding both maxima and minima. Seasonal variation (short time scale mode) of the light curves has been seen from photometry. In addition, long time scale variation is evident through comparison between the present observations and the previous ones Albayrak et al. (2004). Both variations have no regular trend. These fluctuations affect, to some extent, the determination of the system physical and geometrical parameters. The SW Lac system exhibits variable O'Connell effect as seen from Table (3). This effect and its variation can be well explained by starspot hypothesis on the components of W UMa systems that posses high surface activity because of their rapid rotation and convective envelopes.

New nine times of minima of SW Lac. (Five primaries and four secondaries) were derived from the present photometry. A new linear ephemeris was determined as: Hel. J.D. (Min.I) =2449982.5347 (±0.000141) + 0.320854 (±0.000134)*E

The visual and blue solutions provide almost similar values for orbital inclination and the components' physical parameters (mass, radius and surface temperature). These parameters are, in principle, agreed with those derived by others (e.g. Jeong et al. 1984).

The solutions reveal that SW Lac is an overcontact close binary system by 31-37 %, a shallow common convective envelope that may be responsible for its high activity.

The major differences found in the present solutions are only the spotted areas' parameters. On the one hand, the solution shows that about 2% of the visible surface of cool star has temperature of about 4000 K (0.75 of the stars temperature). In view of solar activity (e.g. Sunspot), both single and double spots aspects can be accepted. Nevertheless, the latter is more appropriate for SW Lac modeling. Where, SW Lac is one of strong chromospheric active W UMa systems that have been found to be correlated with its light curves as well as the phase of observations (Eaton 1983).

# REFERENCES

Albayrak, B., Djurasevic, G., Erkapic, S. & Tanriverdi, T., 2004, A&A, 420, 1039

Barbera, R., 1996, AVE (Analisis de Variabilidad Estelar) Version 2.5, http://www.gea.cesca.es

Bradstreet, D. H. and Steelman, D. P., 2004, Binary Maker 3.0 User Manual.

Eaton, J.A., 1983, ApJ, 268, 800.

Faulkner, D. R. and Bookmyer, B., 1980, PASP, 92, 92.

Jeong, J. H.; Kang, Y. W.; Lee, W. B.; and Sung, E. C., 1994, ApJ, 421, 779-786.

Johnson, H. L. & Morgan, W. W., 1953, ApJ, 117, 313.

Kwee, K. K. and Van Woerden, H., 1956, Bull. Astron. Inst. Neth. 12, 327.

Leavitt, H.S., 1918, Harvard Obs. Circ., No. 207.

Leung, K. C.; Zhai, D.; and Zhang, R., 1984, PASP, 96, 634.

Niarchos, P.G., 1987, Astron. Astrophys. Supppl. Ser., 67, 365-371.

Rucinski, S.M., 1968, Acta Astron., 18, 49.

Van Hamme, W., & Wilson, R. E., 2003, ASP Conf. Ser. 298, GAIA Spectroscopy, Science and Technology, ed. U. Munari (San Francisco: ASP), 323

Wilson, R. E., 1979, ApJ, 234, 1054

Wilson, R. E., 1990, ApJ, 356, 613

Wilson, R. E., 1993, in ASP Conf. Ser. 38, New Frontiers in Binary Star Research, ed. K.C.

Wilson, R. E., & Devinney, E. J., 1971, ApJ, 166, 605